\documentclass[manuscript,screen,nonacm]{acmart}
\AtBeginDocument{%
  }
    
\usepackage{amsmath,amsfonts}
\usepackage{algorithm}
\usepackage{algpseudocode} 
\usepackage{array}
\usepackage{subfig}
\usepackage{textcomp}
\usepackage{stfloats}
\usepackage{url}
\usepackage{verbatim}
\usepackage{graphicx}
\usepackage{multirow}
\usepackage{pifont}
\usepackage{hyperref}
\usepackage{wasysym}
\usepackage{upgreek}

\begin{document}

\title{QFOR: A Fidelity-aware Orchestrator for Quantum Computing Environments using Deep Reinforcement Learning}
\author{Hoa T. Nguyen}
\email{thanhhoan@student.unimelb.edu.au}
\orcid{0000-0001-6904-6312}
\affiliation{%
  \institution{Quantum Cloud Computing and Distributed Systems (qCLOUDS) Laboratory, School of Computing and Information Systems, The University of Melbourne}
  \city{Parkville}
  \state{Victoria}
  \country{Australia}}

\author{Muhammad Usman}
\orcid{0000-0003-3476-2348}
\affiliation{%
  \institution{School of Physics, The University of Melbourne}
  \city{Parkville}
  \state{Victoria}
  \country{Australia}}
  \affiliation{%
  \institution{Data61, CSIRO}
  \city{Clayton}
  \state{Victoria}
  \country{Australia}}
\email{musman@unimelb.edu.au}

\author{Rajkumar Buyya}
\orcid{0000-0001-9754-6496}
\affiliation{%
  \institution{Quantum Cloud Computing and Distributed Systems (qCLOUDS) Laboratory, School of Computing and Information Systems, The University of Melbourne}
  \city{Parkville}
  \state{Victoria}
  \country{Australia}}
\email{rbuyya@unimelb.edu.au}

\renewcommand{\shortauthors}{H. Nguyen, M. Usman, and R. Buyya}




\begin{abstract}
Quantum cloud computing enables remote access to quantum processors, yet the heterogeneity and noise of available quantum hardware create significant challenges for efficient resource orchestration. These issues complicate the optimization of quantum task allocation and scheduling, as existing heuristic methods fall short in adapting to dynamic conditions or effectively balancing execution fidelity and time. Here, we propose QFOR, a \textbf{Q}uantum \textbf{F}idelity-aware \textbf{O}rchestration of tasks across heterogeneous quantum nodes in cloud-based environments using Deep \textbf{R}einforcement learning. We model the quantum task orchestration as a Markov Decision Process and employ the Proximal Policy Optimization algorithm to learn adaptive scheduling policies, using IBM quantum processor calibration data for noise-aware performance estimation. Our configurable framework balances overall quantum task execution fidelity and time, enabling adaptation to different operational priorities. Extensive evaluation demonstrates that QFOR is adaptive and achieves significant performance with 29.5-84\% improvements in relative fidelity performance over heuristic baselines. Furthermore, it maintains comparable quantum execution times, contributing to cost-efficient use of quantum computation resources.

\end{abstract}



\maketitle

\section{Introduction}
The rapid advancement of quantum computing promises to solve computationally intractable problems across critical technology domains, including cryptography \cite{portmann_securityquantumcryptography_2022}, drug discovery \cite{zinner_quantumcomputingspotential_2021}, optimization \cite{lotshaw_scalingquantumapproximate_2022}, and machine learning \cite{biamonte_quantummachinelearning_2017}. 
However, given the significant challenges associated with operating physical quantum hardware, such as stringent environmental requirements and high costs, Quantum Cloud Computing (QCC) has emerged as an access paradigm \cite{nguyen_quantumcloudcomputing_2024}. QCC platforms, offered by major providers like IBM Quantum, Amazon Braket, Google Cloud, and Microsoft Azure, democratize access to Quantum Processing Units (QPUs) in a Quantum-as-a-Service (QaaS) model. This allows users to execute quantum applications remotely without the need for on-premise hardware. 

\begin{figure}[htbp]
\centering
\includegraphics[width=4.8in]{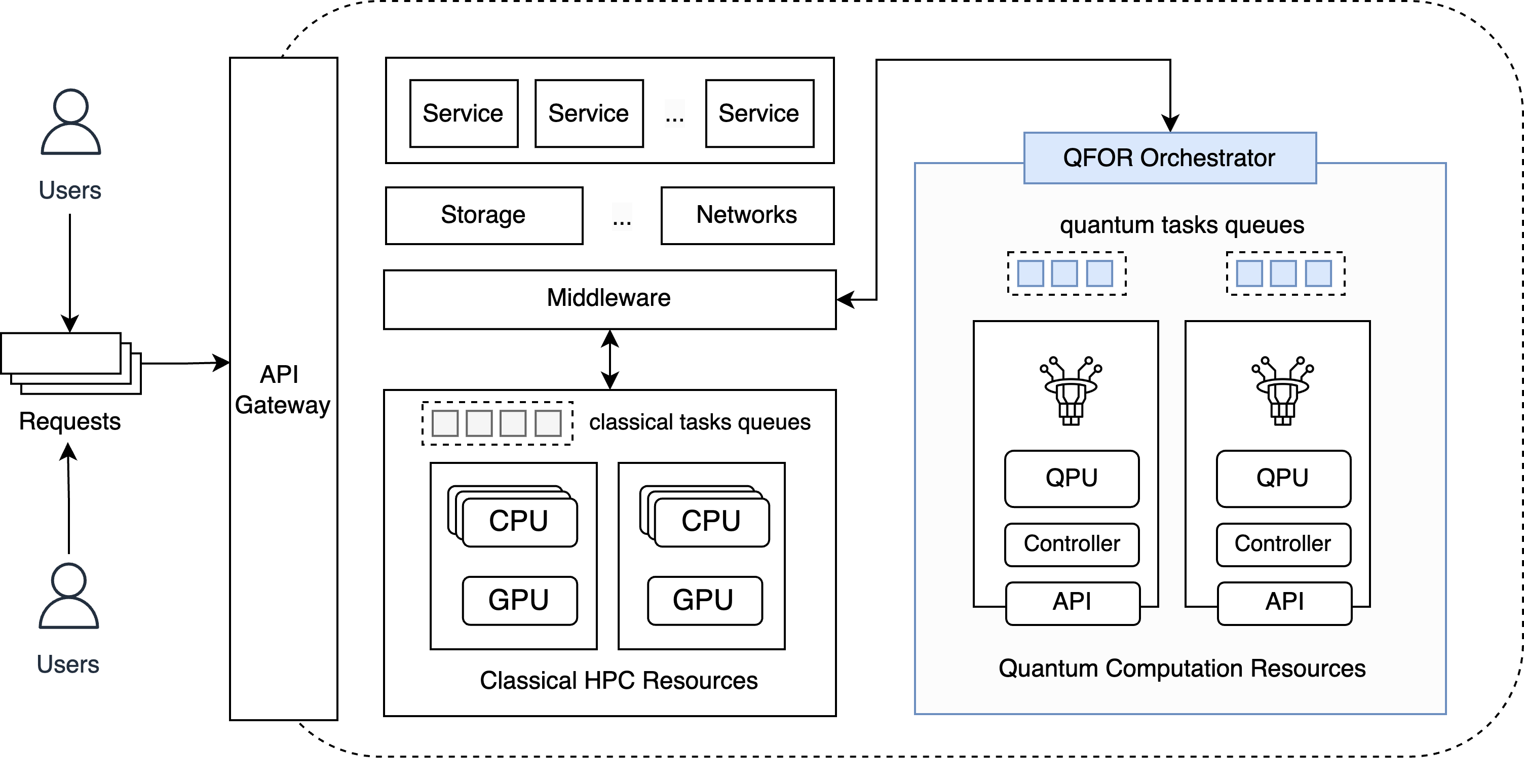}
\caption{High-level view of a Quantum-HPC system in the cloud-based environment. The QFOR Orchestrator (proposed in this work) manages quantum task scheduling and coordination across quantum processing units (QPUs).}
\label{fig_qcloud}
\end{figure}

Notably, current QPUs are not fully fault-tolerant processors and operate within the constraints of the Noisy Intermediate-Scale Quantum (NISQ) era \cite{preskill_quantumcomputingnisq_2018}, and executing quantum applications on them is a hybrid process, involving both classical and quantum execution. For instance, Variational Quantum Algorithms (VQAs) necessitate a classical optimisation step for convergence \cite{moll_quantumoptimizationusing_2018}. Even fully quantum algorithms require classical transpilation to be compatible with the qubit topology and native gate set of the targeted QPU. Furthermore, realizing quantum advantage requires seamless integration with classical high-performance computing (HPC) infrastructure to form hybrid quantum-classical systems \cite{saurabh_conceptualarchitecturequantumhpc_2023, beck_integratingquantumcomputing_2024}. These hybrid systems leverage classical HPC resources for pre-processing, optimization, and post-processing while utilizing quantum processing units (QPUs) for quantum-specific computations. Figure~\ref{fig_qcloud} presents a high-level architecture of a quantum-classical hybrid cloud system, wherein user requests are routed via an API gateway to a corresponding service or application deployment at the middleware layer. Classical tasks for pre-processing and post-processing are queued and dispatched to CPUs and GPUs, while quantum tasks are orchestrated by a quantum task orchestrator component (for example, our proposed QFOR orchestrator in this work), which manages quantum processing units (QPUs) through dedicated queues and APIs.

In these hybrid quantum cloud computing environments, effective quantum task orchestration is crucial for three key reasons. First, quantum computation resources exhibit extreme scarcity as quantum hardware development is still in its early stages \cite{nguyen_quantumcloudcomputing_2024}. Second, quantum resources incur costs significantly higher than equivalent classical compute time, making efficient utilization economically critical. For example, each minute of quantum execution on IBM Quantum hardware costs \$96 USD\footnote{https://www.ibm.com/quantum/pricing}, while each hour of IonQ device reservation on Amazon Braket costs \$7,000 USD\footnote{https://aws.amazon.com/braket/pricing/} (as of June 2025).  Third, quantum fidelity degradation due to poor scheduling can invalidate entire hybrid computations, regardless of classical processing quality \cite{kurlej_performancealgorithmsemerging_2024, giortamis_orchestratingquantumcloud_2024}.
Current quantum hardware exhibits significant variability across multiple dimensions that directly impact quantum task execution \cite{nguyen_quantumcloudcomputing_2024}. Multiple QPU technologies exist, including superconducting qubits, trapped ions, and neutral atoms, each with distinct performance and noise characteristics. Even within the same technology, different QPU models possess varied qubit layouts, native gate sets, and qubit connectivity. Architecturally, devices differ in qubit connectivity constraints how quantum circuits can be efficiently mapped and executed. Additionally, gate durations and error rates vary not only between devices but also over time due to calibration cycles and environmental factors \cite{chen_benchmarkingtrappedionquantum_2023, resch_benchmarkingquantumcomputers_2022}. Device availability is another critical consideration, as queue times can fluctuate based on user demand and system maintenance \cite{ravi_adaptivejobresource_2021}. These factors collectively introduce substantial heterogeneity and uncertainty into the quantum cloud environment. As a result, effective orchestration must be both fidelity-aware, dynamically select quantum computation nodes that optimize for execution reliability while maintaining efficiency with reasonable execution time. This orchestration approach is essential for harnessing the potential of quantum cloud resources and ensuring consistent, high-quality results and cost efficiency for users.

Despite the increasing interest in QCC, existing research in quantum cloud resource management exhibits several critical gaps, particularly concerning heterogeneous quantum computation resources and noise-aware execution fidelity \cite{nguyen_quantumcloudcomputing_2024}. Traditional resource management approaches fall into two categories, both with critical limitations. Traditional HPC scheduling algorithms excel at managing classical computation resources but can struggle to account for quantum-specific constraints such as fidelity optimization, decoherence, and calibration-dependent performance \cite{esposito_hybridclassicalquantumhpc_2023, beck_integratingquantumcomputing_2024}. Conversely, quantum-specific heuristics \cite{ravi_adaptivejobresource_2021, li_qusplitachievingboth_2025, nguyen_iquantumtoolkitmodeling_2024, nguyen_qfaasserverlessfunctionasaservice_2024} are inherently limited in their ability to adapt to the dynamic and uncertain nature of quantum cloud environments. Recent AI-driven approaches, such as deep reinforcement learning (DRL), have shown promise for addressing this problem with several successful cases in the classical cloud-edge \cite{goudarzi_distributeddeepreinforcement_2021, fan_drasdeepreinforcement_2022} and high-performance computing domain \cite{yi_efficientcomputeintensivejob_2020, zhang_robustnessanalysisenhancement_2022}, as it is well-suited for sequential decision-making in environments with incomplete information and stochastic dynamics. However, DRL approaches for quantum cloud orchestration remain limited in scope. Existing works mainly focused on completion time \cite{nguyen_drlqdeepreinforcement_2024} and device allocation \cite{li_moiraioptimizingquantum_2024} or overall fidelity of the targeted quantum device \cite{luo_adaptivejobscheduling_2025}. To our knowledge, no existing work considers the trade-off between circuit execution fidelity, time, and complexity of quantum tasks using DRL-based approaches with real quantum circuit workload. 

Furthermore, as quantum cloud resources are scarce and limited, designing and evaluating resource management in a practical environment is extremely challenging. Therefore, simulation frameworks \cite{nguyen_iquantumtoolkitmodeling_2024, nguyen_qsimpylearningcentricsimulation_2025, luo_adaptivejobscheduling_2025} for modeling and simulating quantum cloud computing environments are essential for this research area. However, existing approaches mainly focus on high-level metrics for the performance estimation of quantum tasks, such as quantum volume \cite{cross_validatingquantumcomputers_2019} and circuit layer operation per second (CLOPS) \cite{wack_qualityspeedscale_2021}. Although this approach is promising, it lacks the adaptability to different structures of quantum circuits and their transpilation to different qubit topologies of quantum hardware. Indeed, specific information on gate errors and durations of individual qubits in quantum hardware can be used to enhance the estimation process. Thus, there is a clear need for resource estimation that systematically explores this fidelity-runtime tradeoff to aid optimal quantum resource orchestration. 

To address the above-mentioned challenges, we propose QFOR, a \textbf{Q}uantum \textbf{F}idelity-aware \textbf{O}rchestrator using a Deep \textbf{R}einforcement learning-based approach that optimizes the overall performance of task orchestration in quantum cloud environments. QFOR models quantum task placement as a Markov Decision Process and employs Proximal Policy Optimization (PPO) \cite{schulman_proximalpolicyoptimization_2017} to learn adaptive policies that balance overall execution fidelity and time of quantum tasks. 

The major contributions and novelty of our work are:
\begin{itemize}
    \item We propose a novel deep reinforcement learning-based task orchestration framework for quantum computing in cloud-based environments. Our method considers the critical trade-off between execution fidelity and quantum execution time, with a primary focus on maximizing fidelity-aware overall performance.
    \item Our approach employs a systematic quantum task execution and fidelity estimators, and mimics the execution of noisy devices based on calibration data and actual quantum circuit properties, enabling more rigorous quantum cloud modeling and simulation of quantum hardware behavior and improved orchestration decisions.
    \item We provide configurable orchestration objectives that balance execution fidelity and latency, accounting for circuit complexity and task priority. Extensive evaluation demonstrates that QFOR is flexible and adaptive, achieving 29.5–84\% improvements in relative fidelity compared to heuristic baselines, while maintaining comparable quantum execution times, thus supporting cost-efficient quantum resource usage.
\end{itemize}

The rest of the paper is organized as follows: Section \ref{sec:related-works} reviews existing works related to our study. Section \ref{sec:system-model} describes the system model and formulates the problems of quantum task orchestration in heterogeneous quantum cloud computing environments. We provide details of the methodology and design of the QFOR framework in Section \ref{sec:QFOR-framework}. Then, Section \ref{sec:evaluation} describes the evaluation study of our proposed framework, followed by and further discussion on the results. Finally, we conclude the paper with key insights, limitations, and future directions in Section \ref{sec:conclusions}.

\section{Related Work}
\label{sec:related-works}
In this section, we review existing literature on quantum resource management and orchestration, categorizing approaches by their methodology and highlighting critical gaps that our work addresses. Table \ref{tab:related-work} provides a comprehensive comparison across key technical dimensions.

\begin{table}[htbp]
\caption{Overall comparison of related works on quantum cloud task orchestration and resource management problem.\\  \small{\textit{(H: Heuristic, DRL: Deep Reinforcement Learning, N/A: Not available, $\LEFTcircle$: Partially Addressed/Described, Env: Environments, E: Emulation with real quantum circuit compilation and execution, S: Simulation with circuit feature or synthetic data)}}}
\label{tab:related-work}
\begin{tabular}{|l|ll|ll|l|lll|l|}
\hline
\multirow{2}{*}{\textbf{Works}}     & \multicolumn{2}{c|}{\textbf{Quantum Tasks}}                                                                                   & \multicolumn{2}{c|}{\textbf{Quantum Nodes}}                                                                    & \multirow{2}{*}{\textbf{\begin{tabular}[c]{@{}l@{}}Method\end{tabular}}} & \multicolumn{3}{l|}{\textbf{Optimization config}}                                                  & \multirow{2}{*}{\textbf{\begin{tabular}[c]{@{}l@{}}Env\\ \end{tabular}}} \\ \cline{2-5} \cline{7-9}
 & \multicolumn{1}{l|}{\textit{\begin{tabular}[c]{@{}l@{}}Real \\ Circuit\end{tabular}}}  & \textit{\begin{tabular}[c]{@{}l@{}}Dataset/Task\\ Generator\end{tabular}} & \multicolumn{1}{l|}{\textit{\begin{tabular}[c]{@{}l@{}}Noise \\ Aware\end{tabular}}} & \textit{Qubit}  &                                                                                 & \multicolumn{1}{l|}{\textit{Fidelity}}  & \multicolumn{1}{l|}{\textit{Time}}      & \textit{Weighted}  &                                                                                            \\ \hline
QFaaS \cite{nguyen_qfaasserverlessfunctionasaservice_2024}                                                                                                         & \multicolumn{1}{l|}{\ding{51}}                                                                 & Qiskit                                                             & \multicolumn{1}{l|}{\ding{51}}                                                       & 7-127           & H                                                                               & \multicolumn{1}{l|}{\LEFTcircle}                 & \multicolumn{1}{l|}{\LEFTcircle}                 & \ding{55}          & E                                                                                        \\ \hline
Ravi et al. \cite{ravi_adaptivejobresource_2021}                                                                                                  & \multicolumn{1}{l|}{\ding{51}}       & Qiskit                                                             & \multicolumn{1}{l|}{\ding{51}}                                                       & 1-65            & H                                                                               & \multicolumn{1}{l|}{\ding{51}}          & \multicolumn{1}{l|}{\LEFTcircle}                 & \ding{55}          & S                                                                                 \\ \hline
iQuantum \cite{nguyen_iquantumtoolkitmodeling_2024}                                                                                                     & \multicolumn{1}{l|}{\ding{51}}                                                                 & MQTBench                                                           & \multicolumn{1}{l|}{\ding{55}}                                                       & 27-127          & H                                                                               & \multicolumn{1}{l|}{\ding{55}}          & \multicolumn{1}{l|}{\ding{51}}                 & \ding{55}          & S                                                                                 \\ \hline
Qonductor \cite{giortamis_orchestratingquantumcloud_2024}                                                                                                 & \multicolumn{1}{l|}{\ding{51}}                                                          & MQTBench                                                           & \multicolumn{1}{l|}{\ding{51}}                                                       & 27              & H                                                                               & \multicolumn{1}{l|}{\ding{51}}          & \multicolumn{1}{l|}{\LEFTcircle}          & \ding{55}          & E                                                                                        \\ \hline
QuSplit \cite{li_qusplitachievingboth_2025}                                                                                                   & \multicolumn{1}{l|}{\ding{51}}                                                                & Qiskit (VQE)                                                       & \multicolumn{1}{l|}{\ding{51}}                                                       & N/A             & H                                                                               & \multicolumn{1}{l|}{\ding{51}}          & \multicolumn{1}{l|}{\ding{55}}          & \ding{55}          & S                                                                                 \\ \hline
DRLQ \cite{nguyen_drlqdeepreinforcement_2024}                                                                                                       & \multicolumn{1}{l|}{\ding{51}}                                                                & MQTBench                                                           & \multicolumn{1}{l|}{\ding{55}}                                                       & 27-127          & DRL                                                                             & \multicolumn{1}{l|}{\ding{55}}          & \multicolumn{1}{l|}{\ding{51}}          & \ding{55}          & S                                                                                 \\ \hline
Moirai \cite{li_moiraioptimizingquantum_2024}                                                                                                      & \multicolumn{1}{l|}{\ding{51}}        & Qiskit                                                             & \multicolumn{1}{l|}{\ding{55}}                                                       & 5-7             & DRL                                                                             & \multicolumn{1}{l|}{\ding{55}}          & \multicolumn{1}{l|}{\ding{51}}          & \ding{55}          & E                                                                                        \\ \hline
Luo et al. \cite{luo_adaptivejobscheduling_2025}                                                                                                    & \multicolumn{1}{l|}{\ding{55}}                                                            & Random data                                                          & \multicolumn{1}{l|}{\ding{51}}                                                       & 127             & DRL                                                                             & \multicolumn{1}{l|}{\ding{51}}          & \multicolumn{1}{l|}{\LEFTcircle}          & \ding{55}          & S                                                                                 \\ \hline
\textit{\textbf{QFOR}}                                                                 & \multicolumn{1}{l|}{\textbf{\ding{51}}}                                              & \textbf{MQTBench}                                                  & \multicolumn{1}{l|}{\textbf{\ding{51}}}                                              & \textbf{27-127} & \textbf{DRL}                                                                    & \multicolumn{1}{l|}{\textbf{\ding{51}}} & \multicolumn{1}{l|}{\textbf{\ding{51}}} & \textbf{\ding{51}} & \textbf{E}                                                                         \\ \hline
\end{tabular}
\end{table}

Early papers in quantum cloud orchestration have primarily relied on heuristic approaches that extend classical scheduling paradigms to quantum environments. Ravi et al. \cite{ravi_adaptivejobresource_2021} introduced an adaptive job and resource management for quantum clouds, focusing on fidelity optimization through device selection. However, their approach relies on predetermined heuristics and statistical analysis that can struggle to adapt to temporal variations in quantum device performance. QFaaS \cite{nguyen_qfaasserverlessfunctionasaservice_2024} is one of the first serverless quantum computing framework, establishing a Function-as-a-Service model for quantum task execution. While QFaaS demonstrates practical hybrid quantum-classical integration across multiple cloud providers, it employs primarily heuristic models for resource allocation decisions based on execution priority (speed and accuracy) using Quantum Volume (QV) \cite{cross_validatingquantumcomputers_2019}, CLOPS \cite{wack_qualityspeedscale_2021}, and queue metrics without considering dynamic fidelity-runtime tradeoffs. Pioneering the discrete-event quantum cloud modeling and simulation, iQuantum \cite{nguyen_iquantumtoolkitmodeling_2024} and QSimPy \cite{nguyen_qsimpylearningcentricsimulation_2025} provide comprehensive toolkits for quantum cloud resource management design and evaluation, but have not fully considered noise-aware modeling. This gap is also one of the key motivations for QFOR to address, providing an emulation approach to mimic the noisy-quantum hardware in practical quantum cloud environments. Qonductor \cite{giortamis_orchestratingquantumcloud_2024} represents one of the most sophisticated heuristic frameworks, offering a hybrid quantum-classical orchestration. Despite introducing important concepts like hybrid resource estimation, Qonductor relies on heuristic scheduling policies and a prediction model based on historical data that can be challenging to adapt to dynamic quantum environments. Focusing only on quantum optimization applications, QuSplit \cite{li_qusplitachievingboth_2025} focuses on optimizing fidelity and throughput through job splitting using a genetic algorithm. However, the limitation of heuristic approaches lies in their ability to adapt to the dynamic and stochastic nature of quantum cloud environments. As quantum hardware and the characteristics of quantum tasks evolve, static scheduling policies become harder to adapt to, requiring adaptive learning-based approaches.

Recent research has begun exploring deep reinforcement learning as a solution to quantum orchestration challenges, demonstrating significant improvements over heuristic baselines while revealing important limitations. DRLQ \cite{nguyen_drlqdeepreinforcement_2024} pioneered the use of deep reinforcement learning \cite{hessel_rainbowcombiningimprovements_2017} for quantum task scheduling, demonstrating significant improvements over heuristic baselines in terms of completion time. Similarly, Moirai \cite{li_moiraioptimizingquantum_2024} employed policy gradient methods within the OpenWhisk framework to schedule quantum circuits on small-scale quantum devices. However, neither DRLQ nor Moirai incorporated comprehensive noise-awareness in the decision-making process. Recently, Luo et al. \cite{luo_adaptivejobscheduling_2025} employed a DRL approach in a simulated environment to maximize targeted device fidelity on different 127-qubit quantum nodes. However, their work relied on CLOPS-based estimation for quantum execution with synthetic random data for the training and evaluation, rather than real quantum circuits, and did not consider the execution time factor in the DRL policy design to fully address the tradeoff between time and fidelity of task execution.

As summarized in Table \ref{tab:related-work}, our work addresses several limitations in the existing approaches. 
Existing approaches lack comprehensive performance modeling that integrates circuit features, device calibration data, and noise-aware fidelity and execution time estimation for improving scheduling decisions. Besides, current DRL-based methods operate on limited-scale systems or synthetic random data in simulated environments rather than emulating the execution with quantum circuit compilation and execution, which have limitations in practical applicability to quantum cloud environments. Our work provides a comprehensive, deep reinforcement learning-based framework that considers noise-aware performance modeling using device calibration data and configurable optimization with evaluation on realistic quantum circuit workloads from a well-known quantum circuit dataset (MQTBench \cite{quetschlich_mqtbenchbenchmarking_2023}). Our work aims to contribute valuable approaches and insights for future works in quantum cloud orchestration, enabling learning-driven resource management that adapts to the dynamic, heterogeneous nature of NISQ-era quantum computing environments while optimizing for practical deployment requirements.

\section{System Model and Problem Formulation}
\label{sec:system-model}
This section presents our system model and problem formulation for quantum cloud orchestration, focusing on fidelity-aware performance metrics.

\subsection{Quantum Task Model}
In quantum cloud environments, a quantum task (QTask) represents the unit of quantum computation requiring adaptive resource orchestration to optimise the performance. A QTask can comprise one or multiple quantum circuits that need to be executed with specific qubit requirements, gate operations, and circuit depth. In the context of this work, we consider each QTask as a single independent quantum circuit. These QTasks can encompass different quantum algorithms, each with distinct characteristics. 

Let $\Gamma = \{\tau_1, \tau_2, \ldots, \tau_N\}$ denote a sequence of quantum tasks arriving for execution. Each task $\tau_i \in \Gamma$ is characterized by the following properties:
\begin{equation}
\tau_i = \left( a_i, q_i, d_i, s_i, g^{(1)}_i, g^{(2)}_i, \mathcal{C}_i \right)
\end{equation}
where:
\begin{itemize}
    \item $a_i \in \mathbb{R}_{\geq 0}$: arrival time of the task into the system.
    \item $q_i \in \mathbb{N}$: number of qubits required for execution.
    \item $d_i \in \mathbb{N}$: circuit depth, i.e., the longest gate dependency path.
    \item $s_i \in \mathbb{N}$: number of shots (i.e., repetition of the execution)
    \item $g^{(1)}_i \in \mathbb{N}$: total number of single-qubit gates.
    \item $g^{(2)}_i \in \mathbb{N}$: total number of two-qubit gates.
    \item $\mathcal{C}_i$: the quantum circuit representation as a Directed Acyclic Graph (DAG).
\end{itemize}

Each circuit $\mathcal{C}_i$ is modeled as a DAG $\mathcal{G}_i = (\mathcal{V}_i, \mathcal{E}_i)$, where $\mathcal{V}_i$ is the set of all quantum operations (gates), and $\mathcal{E}_i \subseteq \mathcal{V}_i \times \mathcal{V}_i$ represents directed edges indicating gate dependencies.

\begin{figure}[htbp]
\centering
\includegraphics[width=5in]{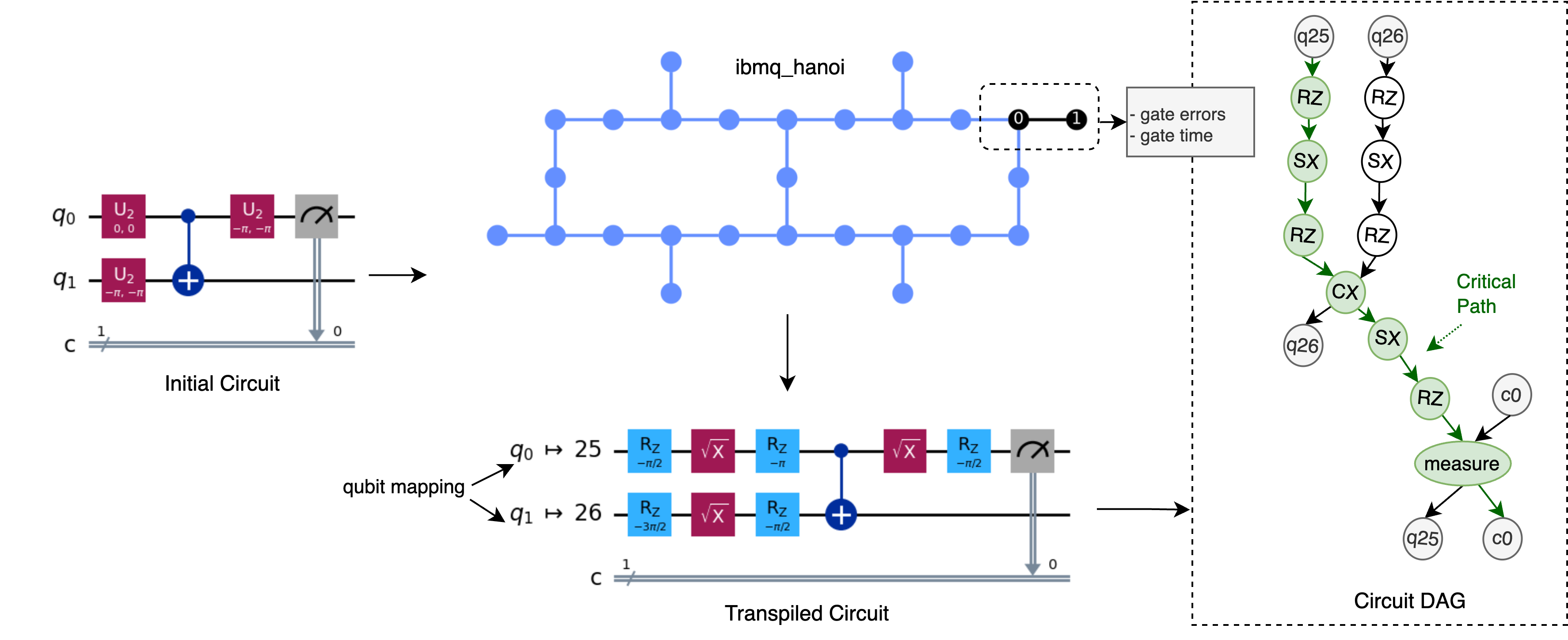}
\caption{Example of an initial quantum circuit and its transpilation, qubit mapping to a 27-qubit QNode (ibmq\_hanoi), and DAG of the transpiled circuit with the critical path illustration. }
\label{fig_estimator}
\end{figure}

Example of a quantum circuit within a QTask and its transpilation, mapping and DAG representation are illustrated in Figure \ref{fig_estimator}. The DAG representation enables critical path analysis for the estimation of QTask execution time and the execution fidelity based on the calibration data of QNodes. Based on the current state of typical quantum cloud environments \cite{nguyen_quantumcloudcomputing_2024}, we assume each task requires exclusive access to a set of qubits on a quantum processor and no preemption or interruption occurs during task execution. The orchestration focuses exclusively on quantum circuit execution, which dominates in terms of resource sensitivity compared to classical resources.

\subsection{Quantum Computation Resource Model}
The quantum cloud infrastructure consists of heterogeneous quantum nodes (QNodes) with one or multiple quantum processing units (QPUs) with distinct performance characteristics that directly impact scheduling decisions. In the context of this work, we consider a single QPU per QNode to reflect the current state of available quantum hardware. Each QNode exhibits unique hardware specifications and properties, which must be taken into account during scheduling to achieve reliable and efficient quantum task execution.

Let the set of quantum resources be denoted by: $\mathcal{N} = \{ n_1, n_2, \dots, n_M \}$, where each node $n_j \in \mathcal{N}$ is defined as:
\begin{equation}
n_j = \left( q_j, \mathcal{G}_j, \mathcal{D}_j, \mathcal{E}_j, \rho_j(t) \right)
\end{equation}
where:
\begin{itemize}
    \item $q_j = |\mathcal{Q}_j| \in \mathbb{N}$: number of physical qubits and $\mathcal{Q}_j$ is the set of all available qubits at QNode $n_j$.
    \item $\mathcal{G}_j = (\mathcal{V}_j, \mathcal{E}_j)$ is the qubit connectivity graph, where $V_j$ is the set of all qubits and $E_j$ is the set of all edges connecting these qubits.
    \item $\mathcal{D}_j: \mathcal{O}_j \times \mathcal{Q}_j \rightarrow \mathbb{R}_{>0}$ maps available gates $\mathcal{O}_j$ and qubits $\mathcal{Q}_j$ to corresponding gate execution durations
    \item $\mathcal{E}_j: \mathcal{O}_j \times \mathcal{Q}_j \rightarrow [0,1]$ maps all available gates $\mathcal{O}_j$ and qubits $\mathcal{Q}_j$ to corresponding error probabilities.
    \item $\rho_j(t)$: other dynamic state of the QNode at time t, such as next available time and queuing information.
\end{itemize}

This abstraction allows us to capture both static and time-dependent dynamics of each quantum node. These features are critical for orchestrator decisions in dynamic workload settings and heterogeneous quantum cloud environments.
\subsection{Fidelity-aware Orchestration Performance Model}
For quantum cloud orchestration in the NISQ era, there are two critical metrics that indicate the performance of the orchestration decision: execution fidelity and time. First, execution fidelity can be considered as one of the most critical factors. Quantum execution exhibits high sensitivity to noise, where small fidelity degradations can render results meaningless regardless of execution speed \cite{wack_qualityspeedscale_2021}. Poor device selection as well as poor selection of qubits on the targeted device can reduce algorithmic success probability, necessitating multiple re-executions that far exceed any time savings from faster scheduling. Second, quantum task execution time in the quantum node is also critical, as quantum resources are scarce and extremely expensive \cite{nguyen_quantumcloudcomputing_2024}. Besides, quantum states decay exponentially with time, making execution time a fundamental physical constraint rather than merely an optimization preference \cite{kordzanganeh_benchmarkingsimulatedphysical_2023}. Longer execution sequences suffer increased error accumulation, creating a direct coupling between execution time and fidelity. Therefore, we define the orchestration performance metric that mainly focuses on the fidelity of the execution, while maintaining the execution time and considering the complexity of the task needs to be executed, as well as how good it compares to other available decisions.

\subsubsection{Fidelity Performance Score}
The execution fidelity performance score $\mathcal{F}_{i,j}$ is the key objective in this work and captures the comprehensive quality of executing task $\tau_i$ on node $n_j$ through three complementary components:

For task $\tau_i$ assigned to node $n_j$, the base execution fidelity $F_{i,j}$ is approximately estimated as:
\begin{equation}
F_{i,j} = \prod_{g \in \mathcal{C}'_i} \left(1 - \mathcal{E}_j(g, \vec{q}_g)\right)
\end{equation}
where $\mathcal{C}'_i$ represents the transpiled circuit and $\vec{q}_g$ are the assigned physical qubits. 

The base fidelity score $R_{rf}$ normalizes this against expected fidelity performance $F^{'}_{i,j}$ based on the average gate errors of all available quantum nodes, preventing hardware-specific biases while rewarding above-expected performance.

\begin{equation}
    R_{rf} = \dfrac{F_{i,j}}{F^{'}_{i,j}}
\end{equation}

In quantum cloud environments, simpler circuits (i.e., those with shallow depth and fewer gates) tend to achieve higher fidelity due to reduced exposure to noise. As a result, an orchestration policy that solely maximizes fidelity may develop a bias toward such circuits, systematically deprioritizing more complex tasks. To address this issue, we introduce a small complexity bonus $R_{cb}$, which encourages the orchestrator to also consider more complex circuits:
\begin{equation}
R_{cb} = w_d \cdot \dfrac{d_i}{D_{max}} + w_g \cdot \sum \dfrac{g_i}{G_{max}}
\end{equation}
where $d_i$ is circuit depth, $\sum g_i$ is total gate count in the circuit of the scheduled QTask, $w_d$ and $w_g$ are adjustable weights, and $D_{max}, G_{max}$ are normalization bounds for the maximum depth and gates of the circuit. This ensures fairness and task diversity within the orchestration policy while remaining computationally efficient and easy to integrate into learning-based decision frameworks. 
We also introduce a ranking bonus $R_{rb}$ that captures the relative quality of selecting a specific quantum node compared to all available options for a given task to ensure that the orchestration policy not only aims for high absolute fidelity but also makes competitively optimal decisions based on the current system state. It is defined as:
\begin{equation}
R_{rb} = \frac{F_{i,j} - F_{worst}}{F_{best} - F_{worst}}
\end{equation}
where $F_{best}$ and $F_{worst}$ represent the highest and lowest achievable fidelity across all nodes for task $\tau_i$.

The comprehensive execution fidelity performance score (or relative fidelity) combines these components:
\begin{equation}
\label{eq:fidelity}
\mathcal{F}_{i,j} = \alpha_1 \cdot R_{rf} + \alpha_2 \cdot R_{cb} + \alpha_3 \cdot R_{rb}
\end{equation}
where $\alpha_1, \alpha_2,$ and $ \alpha_3$ are configurable weights and were set by default, and $\alpha_1 = 0.8, \alpha_2 = \alpha_3 = 0.1$, emphasizing the key focus of our orchestration on execution fidelity while maintaining circuit complexity and decision ranking awareness. By aggregating these complementary components, $\mathcal{F}_{i,j}$ provides a robust fidelity-based performance measure. It supports fair and adaptive orchestration across diverse workloads and heterogeneous quantum hardware, and serves as a principled reward signal in reinforcement learning-based policy optimization of this work.

\subsubsection{Time Penalty Score}
The time penalty score $\mathcal{T}_{i,j}$ captures the temporal cost of quantum task execution, encompassing both waiting and actual quantum execution time. For task $\tau_i$ assigned to node $n_j$, the quantum execution time is estimated by analyzing the critical path of the transpiled circuit:
\begin{equation}
T^{exec}_{i,j} = s_i \cdot \sum_{g \in \text{CP}(\mathcal{C}'_i)} \mathcal{D}_j(g, \vec{q}_g)
\end{equation}
where $\text{CP}(\mathcal{C}'_i)$ represents the critical path (longest execution sequence) of the transpiled circuit, and $\mathcal{D}_j(g, \vec{q}_g)$ is the gate execution duration along the critical path, $s_i$ is the number of shots (execution iterations). An example of the longest path of a quantum circuit is illustrated in Figure \ref{fig_estimator}. The total completion time includes queuing delays of QTask until it can be executed at the targeted QNode $\mathcal{T}_{i,j} = T^{wait}_{i,j} + T^{exec}_{i,j}$ and is normalized based on the maximum completion time bound to enable stable policy training.


\subsection{Problem Formulation}
Given the fidelity performance score $\mathcal{F}_{i,j}$ and time penalty score $\mathcal{T}_{i,j}$ defined above, we formulate the quantum cloud orchestration problem as a sequential decision-making process that maximizes the combined orchestration performance across all quantum tasks.

The quantum cloud orchestration problem can be formally stated as follows: Given a sequence of quantum tasks $\Gamma = \{\tau_1, \tau_2, \ldots, \tau_N\}$ arriving dynamically in a quantum cloud environment with heterogeneous quantum nodes $\mathcal{N} = \{n_1, n_2, \ldots, n_M\}$, find an optimal orchestration policy $\pi: \Gamma \rightarrow \mathcal{N}$ that assigns each task $\tau_i$ to an appropriate quantum node $n_j = \pi(\tau_i)$ to maximize the overall orchestration performance.
For each assignment decision $\pi(\tau_i)$, the overall orchestration performance is quantified by combining the fidelity performance score and time penalty score, with a negative value of time score indicating the secondary goal to minimize the time penalty:

\begin{equation}
\label{score}
\mathcal{P}_{i,\pi(\tau_i)} = \mathcal{F}_{i,\pi(\tau_i)} -\beta \cdot \mathcal{T}_{i,\pi(\tau_i)}
\end{equation}
where $\mathcal{F}_{i,\pi(\tau_i)}$ represents the relative fidelity performance score, $\beta$ is the configurable time penalty weight and $\mathcal{T}_{i,\pi(\tau_i)}$ represents the time penalty, ensuring that higher fidelity and lower execution time both contribute positively to the overall performance score.
The primary optimization objective is to maximize the cumulative orchestration performance across all quantum tasks can be defined as follows:

\begin{equation}
\max_{\pi} \sum_{i=1}^{N} \mathcal{P}_{i,\pi(\tau_i)} = \max_{\pi} \sum_{i=1}^{N} \left[ \mathcal{F}_{i,\pi(\tau_i)} -\beta \cdot \mathcal{T}_{i,\pi(\tau_i)} \right]
\end{equation}
subject to:
\begin{align}
&C1: Size(\pi(\tau_i)) = 1, \forall \pi(\tau_i) \in \{1, \ldots, M\} \\
&C2: q_{\tau_i} \leq q_{\pi(\tau_i)}, \forall i \in \{1, \ldots, N\} \\
&C3: \mathcal{C}_{\tau_i} \hookrightarrow \mathcal{G}_{\pi(\tau_i)}, \forall i \in \{1, \ldots, N\}
\end{align}

\noindent where $N$ is the total number of quantum nodes, $M$ is the total number of tasks that need to be scheduled. $C1$ shows that each QTask will be allocated to exactly one QNode at a time, $C2$ indicates that the number of qubits in the targeted QNode needs to be larger than or equal to the number of qubits required by the allocated QTask, and $C3$ implies that the quantum circuit of the QTask can be mapped to the QNode through the transpilation process. This optimization problem exhibits several challenges. First, quantum node characteristics ($\mathcal{D}_j, \mathcal{E}_j$) vary with calibration cycles, and queue states $\rho_j(t)$ change with task arrivals and completions, requiring adaptive decision-making capabilities. Second, the discrete assignment decisions combined with non-linear relationships between circuit characteristics and performance metrics create a combinatorial optimization problem. Furthermore, the sequential arrival of quantum tasks with unknown future characteristics necessitates an adaptive optimization without complete future information. The tension between fidelity maximization and time minimization requires sophisticated balancing strategies that adapt to different priorities.

Given these complexities, traditional heuristic approaches cannot effectively navigate the dynamic trade-offs inherent in quantum cloud orchestration. Therefore, we propose a deep reinforcement learning approach that models this problem as a Markov Decision Process, enabling the learning of adaptive orchestration policies that can balance fidelity and time objectives with a set of configurable weights that can be adjusted based on the priority of the orchestration. The sequential nature of the decision-making process, combined with the need for adaptive policy learning, makes reinforcement learning particularly well-suited for this orchestration challenge, which is widely used effectively for resource management in the classical computing environments \cite{goudarzi_distributeddeepreinforcement_2021, chen_adaptiveefficientresource_2022, fan_drasdeepreinforcement_2022}.

\section{QFOR Framework and Technique}
\label{sec:QFOR-framework}

\subsection{Main components and Design}
\begin{figure}[htbp]
\centering
\includegraphics[width=6in]{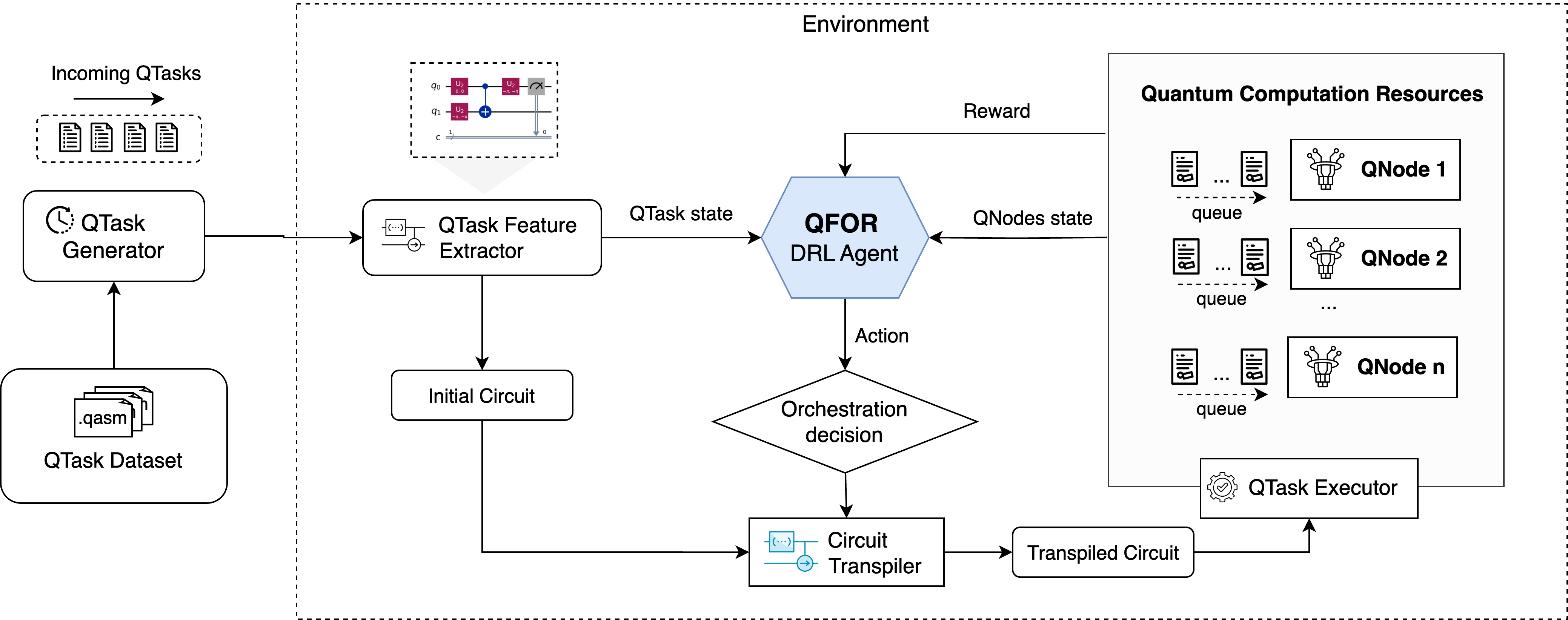}
\caption{Overview of the main components of the QFOR framework}
\label{fig_overview}
\end{figure}

The QFOR framework implements an adaptive orchestration system that applies deep reinforcement learning to optimize quantum task placement in heterogeneous cloud environments. Figure~\ref{fig_overview} illustrates the framework architecture, consisting of six integrated components that collectively enable adaptive fidelity-aware orchestration decisions.

\begin{enumerate}
    \item \textit{QTask Generator and Dataset}: The QTask Generator serves as the workload interface, consuming quantum circuit representations from standardized datasets (for example, MQTBench \cite{quetschlich_mqtbenchbenchmarking_2023}) in OpenQASM format \cite{cross_openqasm3broader_2022}. This component simulates realistic quantum cloud workloads by generating tasks with diverse complexity characteristics—varying qubit requirements, circuit depths, and algorithmic patterns. The generator also supports configurable quantum task arrival patterns to model different cloud workload scenarios.
    \item \textit{QTask Feature Extractor:} This component analyzes quantum circuits to extract critical features, including qubit count, circuit depth, gate statistics (single-qubit, two-qubit gate, measurement counts), and circuit structure. These features are normalized and encoded into a concise representation that characterizes the computational requirements and complexity of each QTask, enabling the DRL agent to make informed scheduling decisions based on circuit characteristics.
    \item \textit{Deep Reinforcement Learning Agent}: The key component that implements the DRL-based decision-making policy. It observes the current state of quantum tasks and available quantum nodes, processes this information through a DRL agent, and produces orchestration decisions that assign tasks to appropriate quantum nodes. The orchestrator continuously learns from execution outcomes through the defined reward function that optimizes the overall performance of the orchestration decision.
    \item \textit{Heterogeneous Quantum Cloud Environment:} The framework extends the capabilities of QSimPy \cite{nguyen_qsimpylearningcentricsimulation_2025} to models of heterogeneous quantum computing resources (QNodes) with varying capabilities, including qubit counts, connectivity graphs, gate durations, and error rates. Each node maintains dynamic state information such as queue length and availability, enabling realistic simulation of quantum cloud environments.
    \item \textit{Circuit Transpiler: }This component transforms logical quantum circuits into hardware-specific implementations optimized for the target quantum node's topology and native gate set. The transpilation process is guided by the orchestration decision and affects both execution time and fidelity. We utilized the Qiskit Transpiler\footnote{https://quantum.cloud.ibm.com/docs/en/api/qiskit/transpiler} with optimisation level 3 as the default transpilation mode. 
    \item \textit{QTask Executor:} This component mimics the execution of a quantum task on noisy quantum nodes by analysing the transpiled quantum circuit and estimating execution fidelity and time metrics based on the system model defined in Section \ref{sec:system-model} and the calibration snapshot data of IBM Quantum systems (with Qiskit Fake Provider\footnote{https://quantum.cloud.ibm.com/docs/en/api/qiskit-ibm-runtime/fake-provider}).
\end{enumerate}

The framework provides a learning system loop where the DRL agent observes environment states, selects quantum node assignments, receives performance feedback through the reward obtained, and iteratively improves scheduling policies. The integration of realistic transpilation, noise-aware execution modeling, and configurable performance objectives enables the system to learn nuanced scheduling strategies that adapt to varying operational priorities while maintaining practical applicability to practical quantum cloud systems.

This modular architecture supports extensibility for different quantum hardware backends, alternative circuit datasets, and enhanced performance models while maintaining the core orchestration capability through learning-based algorithms with design principles similar to existing works \cite{nguyen_iquantumtoolkitmodeling_2024, nguyen_qsimpylearningcentricsimulation_2025}.

\subsection{Deep Reinforcement Learning Model}

Based on the system model and problem formulation in Section~\ref{sec:system-model}, we model the quantum task orchestration as a Markov Decision Process (MDP) to enable adaptive policy training through deep reinforcement learning. The MDP is defined by the tuple $(\mathbb{S}, \mathbb{A}, \mathbb{P}, \mathbb{R}, \gamma)$, where $\mathbb{S}$ is the state space, $\mathbb{A}$ the action space, $\mathbb{P}$ the state transition probability, $\mathbb{R}$ the reward function, and $\gamma \in [0,1]$ the discount factor that balances immediate and future rewards. At each discrete time step $t$, the agent observes the current state $s_t$ of the environment, selects an action $a_t$ according to a policy $\pi(a_t|s_t)$, and moves to the next state $s_{t+1}$ while receiving a reward $r_t$. The objective of the agent is to maximize the expected cumulative discounted reward, defined as $\mathbb{V}^\pi(s_t) = \mathbb{E}_\pi \left[ \sum_{t} \gamma^t r_t\right]$, by learning an optimal policy $\pi$. Typically, the policy is parameterized by a neural network and is improved iteratively through training based on observed transitions and rewards.

\textbf{State Space $\mathbb{S}$:}
The state $s_t$ at time $t \in \mathbb{T}$ is a concatenation of the feature vector of the current quantum task $\tau_i$ and the feature vectors of all available quantum nodes (QNodes) $\mathcal{N}$ in the environment. 
\begin{equation}
s_t = \left( \mathbf{f}^{\tau_i}_t, \mathbf{f}^{\mathcal{N}}_t \right)
\end{equation}
where $\mathbf{f}^{\tau_i}_t \in \mathbb{R}^p$ is the $p$-dimensional feature vector of the current QTask $\tau_i$, and $\mathbf{f}_t^{\mathcal{N}} \in \mathbb{R}^{m \times o}$ is the concatenated feature matrix of all $m$ QNodes, each with $o$ features. All features are normalized to $[0,1]$ to ensure stable learning dynamics across heterogeneous scales. Thus, the State space can be defined as:
\begin{equation}
    \mathbb{S} = \{s_t | s_t = \left( \mathbf{f}^{\tau_i}_t, \mathbf{f}^{\mathcal{N}}_t \right), \forall t \in \mathbb{T}\}
\end{equation}

\textbf{Action Space $\mathbb{A}$:}
The action space is discrete and corresponds to the assignment of a suitable QNode for the placement of the current incoming QTask. At each timestep, the agent chooses action $a_t \in \{0, 1, \ldots, m-1\}$, where $m=|\mathcal{N}|$ and $a_t = i$ denotes assigning the task to QNode $n_i \in \mathcal{N}$. Thus,
\begin{equation}
    \mathbb{A} = \mathcal{N}
\end{equation}

\textbf{Reward Function $\mathbb{R}$:}
The reward function directly implements the orchestration performance model defined in Equation \ref{score}. For each assignment decision to allocate QTask $\tau_i$ to QNode $n_j$ at time $t$, the reward $r_t \in \mathbb{R}$ combines the fidelity score $\mathcal{F}_{i,j}$ and time penalty $\mathcal{T}_{i,j}$ with configurable weight $\beta$:

\begin{equation}
\label{eq:reward}
r_t = 
\begin{cases}
    \mathcal{F}_{i,j} - \beta \cdot \mathcal{T}_{i,j} & \text{if } \text{the execution is successful}
    \\ 
    P_{fail} & \text{otherwise}
\end{cases}
\end{equation}

A failure penalty $P_{\text{fail}}$ is applied when task $\tau_i$ cannot be executed on the selected node for any reason, providing negative feedback for infeasible assignments to improve the decision of the reinforcement learning policy.

\subsection{QFOR: The DRL-based Quantum Fidelity-aware Orchestration Policy}

The QFOR technique extends the Proximal Policy Optimization algorithm \cite{schulman_proximalpolicyoptimization_2017} to the quantum cloud orchestration problem, with the overall training workflow shown in Algorithm \ref{alg:QFOR-ppo}. The algorithm operates on the principle of learning an optimal scheduling policy $\pi_\theta: \mathcal{S} \rightarrow \Delta(\mathbb{A})$ that maximizes long-term cumulative reward while maintaining stable learning dynamics through trust region constraints.

\begin{algorithm}[htbp]
\caption{QFOR Training Workflow with Proximal Policy Optimization}
\label{alg:QFOR-ppo}
\begin{algorithmic}[1]
\Require QTask dataset $\mathcal{QD}$, QNodes $\mathcal{N}$, hyperparameters $\Theta$
\Ensure Trained policy $\pi_\theta$

\State \textbf{Initialize:} Policy network $\pi_\theta$, value network $V_\phi$, experience buffer $\mathcal{B}$
\State \textbf{Initialize:} Quantum Cloud Environment in QSimPy (QSimPyEnv)

\For{iteration $k = 1, 2, \ldots, K$}
    \State $\mathcal{B} \leftarrow \emptyset$ \Comment{Clear experience buffer}
    
    \For{worker $w = 1, 2, \ldots, W$} \Comment{Parallel rollouts}
        \State Reset environment: $(s_0, \tau_0) \sim \text{QSimPyEnv}(\mathcal{QD})$
        
        \For{step $t = 0, 1, \ldots, T-1$}
            \State $s_t \leftarrow f_{\text{state}}(\tau_t, \mathcal{N}, t)$ \Comment{Normalized features}
            \State $a_t \sim \pi_\theta(\cdot | s_t)$ \Comment{Sample action from policy}
            
            \State $ \text{ProcessTask}(\tau_t, n_{a_t})$
            
            \State $r_t \leftarrow \text{CaculateReward}()$
            
            \State $s_{t+1}, \tau_{t+1} \leftarrow \text{NextTask}()$ \Comment{Get next QTask}
            \State $\mathcal{B} \leftarrow \mathcal{B} \cup \{(s_t, a_t, r_t, s_{t+1})\}$ \Comment{Store exp.}
            
            \If{episode terminated or $t = T-1$}
                \State \textbf{break}
            \EndIf
        \EndFor
    \EndFor
    
    \State \textbf{Policy Update:}
    \State Compute advantages: $\hat{A}_t = \delta_t + (\gamma \lambda) \delta_{t+1} + \ldots$ 
    \State Compute returns: $\hat{R}_t = \hat{A}_t + V_\phi(s_t)$
    
    \For{epoch $e = 1, 2, \ldots, E$}
        \For{minibatch $\mathcal{B}_m \subset \mathcal{B}$}
            \State $r_t(\theta) \leftarrow \dfrac{\pi_\theta(a_t|s_t)}{\pi_{\theta_{\text{old}}}(a_t|s_t)}$ \Comment{Probability ratio}
            
            \State $L^{\text{CLIP}}(\theta) \leftarrow \mathbb{E}\left[\min(r_t(\theta)\hat{A}_t, \text{clip}(r_t(\theta), 1-\epsilon, 1+\epsilon)\hat{A}_t)\right]$
            
            \State $L^{\text{VF}}(\phi) \leftarrow \mathbb{E}\left[(V_\phi(s_t) - \hat{R}_t)^2\right]$
            
            \State $L^{\text{ENT}}(\theta) \leftarrow \mathbb{E}\left[H(\pi_\theta(\cdot|s_t))\right]$ \Comment{Entropy bonus}
            
            \State $\theta \leftarrow \theta + \alpha_\pi \nabla_\theta (L^{\text{CLIP}}(\theta) + c_1 L^{\text{ENT}}(\theta))$
            \State $\phi \leftarrow \phi - \alpha_v \nabla_\phi L^{\text{VF}}(\phi)$
        \EndFor
    \EndFor
    
    \State $\theta_{\text{old}} \leftarrow \theta$ \Comment{Update old policy parameters}
\EndFor

\State \Return Trained policy $\pi_\theta$
\end{algorithmic}
\end{algorithm}

Initially, the algorithm initializes three components: (i) a parameterized policy network $\pi_\theta$ with parameters $\theta$ that maps states to action probability distributions, (ii) a value function approximator $V_\phi$ with parameters $\phi$ that estimates state values for advantage computation, and (iii) an experience buffer $\mathcal{B}$ for storing trajectory data. 
Besides, the quantum computing environment instantiation involves configuring heterogeneous quantum nodes to mimic the realistic NISQ hardware parameters, including error rates, gate durations, and qubit connectivity constraints. 

The outer training loop iterates over $K$ policy improvement cycles, where each iteration corresponds to one complete policy update using collected experience data. This structure follows the standard PPO algorithm \cite{schulman_proximalpolicyoptimization_2017} and Ray RLlib \cite{liang_rllibabstractionsdistributed_} of alternating between data collection and policy optimization phases. 
The experience buffer is cleared at the beginning of each iteration to ensure on-policy learning. 
Parallel rollout collection across $W$ workers enables efficient data gathering and improved sample diversity. 
Each worker operates independently, reducing correlation between consecutive experiences and enhancing the robustness of gradient estimates during policy updates. Environment reset initializes a new episode by sampling the initial QTask $\tau_0$ from the QASM-based  \cite{cross_openqasm3broader_2022}circuit dataset $\mathcal{QD}$ and computing the corresponding initial state $s_0$. The stochastic nature of task arrival ensures diverse training scenarios and prevents overfitting to specific task sequences. 

The inner episode loop processes quantum tasks sequentially until episode termination at the final timestep $t$ or when the task limit is reached. Each step corresponds to scheduling one incoming quantum task to an available quantum node. 
Action sampling follows the current policy distribution, where $a_t$ represents the selected QNode index. The stochastic policy enables exploration while the learned parameters $\theta$ bias the selection toward high-reward actions based on accumulated experience. Each QTask execution involves a circuit transpilation process and performance metrics estimation on the selected quantum node $n_{a_t}$ based on the calibration data of the quantum device. Then, the reward function combines multiple performance indicators as defined (see Equation \ref{eq:reward}) in the previous section.






State transition involves advancing to the next quantum task in the episode sequence and updating the environment time. The next state $s_{t+1}$ incorporates updated quantum nodes metrics and the new quantum task's characteristics, maintaining the Markov property essential for policy gradient convergence. Experience tuple storage enables subsequent policy optimization through gradient-based updates. Each tuple $(s_t, a_t, r_t, s_{t+1})$ provides a complete transition record necessary for advantage estimation and policy gradient computation. Episode termination logic ensures proper boundary handling when task limits are reached or no additional tasks are available. This prevents infinite episodes while maintaining consistent episode lengths for stable learning dynamics.

For the policy optimization phase, our QFOR technique leverages the standard PPO algorithm \cite{schulman_proximalpolicyoptimization_2017} and adapts to quantum task orchestration. The advantage computation using Generalized Advantage Estimation (GAE) \cite{schulman_highdimensionalcontinuouscontrol_2018}:

$$\hat{A}_t = \sum_{l=0}^{\infty} (\gamma \lambda)^l \delta_{t+l}$$
where $\delta_t = R_t + \gamma V_\phi(s_{t+1}) - V_\phi(s_t)$ represents the temporal difference error. GAE balances bias and variance in advantage estimation through the $\lambda$ parameter. 

During the policy update phase, PPO's clipped surrogate objective function is used to ensure bounded policy updates:

$$L^{\text{CLIP}}(\theta) = \mathbb{E}\left[\min(r_t(\theta)\hat{A}_t, \text{clip}(r_t(\theta), 1-\epsilon, 1+\epsilon)\hat{A}_t)\right]$$
where $r_t(\theta) = \dfrac{\pi_\theta(a_t|s_t)}{\pi_{\theta_{\text{old}}}(a_t|s_t)}$. The value function loss $L^{\text{VF}}(\phi) = \mathbb{E}[(V_\phi(s_t) - \hat{R}_t)^2]$ and entropy bonus $L^{\text{ENT}}(\theta) = \mathbb{E}[H(\pi_\theta(\cdot|s_t))]$ provide additional optimization objectives for stable learning and adequate exploration. The clipped objective function ensures monotonic policy improvement with high probability, while the adaptive reward structure maintains the Markov property essential for convergence and provides scale-invariant rewards across diverse circuit complexities, enabling consistent learning signals.

\section{Performance Evaluation}
\label{sec:evaluation}
\subsection{Environment Setup}
We use the QSimPy framework \cite{nguyen_qsimpylearningcentricsimulation_2025} for simulation of quantum cloud environments. We also extended it further to support the modeling of noisy quantum nodes using device calibration data and mimic the practical execution process of a quantum task, which comprises the circuit transpilation to selected QNodes before execution. This approach allows us to emulate heterogeneous quantum cloud computing environments and quantum task execution more comprehensively compared to other existing work \cite{luo_adaptivejobscheduling_2025, nguyen_iquantumtoolkitmodeling_2024}. For quantum cloud computation resources, we created a cluster of 5 different quantum nodes ranging from 27 to 127 qubits using the calibration data of IBM devices (using Qiskit FakeBackend instances\footnote{https://quantum.cloud.ibm.com/docs/en/api/qiskit-ibm-runtime/fake-provider}), including \texttt{ibm\_auckland}, \texttt{ibm\_hanoi}, \texttt{ibm\_kolkata}, \texttt{ibm\_brisbane}, and \texttt{ibm\_sherbrooke}. 



For quantum tasks, we created different training and evaluation datasets with 16 different quantum benchmark algorithms with qubit numbers ranging from 2 to 27 and initial circuit depths (before transpilation) of 3–30 layers, derived from the MQT Bench dataset \cite{quetschlich_mqtbenchbenchmarking_2023}. The QTask arrival times were generated following a Poisson distribution, similar to other works \cite{bruneo_stochasticmodelinvestigate_2014, yoon_adaptivedatacenter_2017} to mimic the task arrival at the cloud data center. We use Gymnasium to wrap the QSimPy-based environment and use Ray RLlib \cite{liang_rllibabstractionsdistributed_} for implementing the reinforcement learning training and using Ray Tune \cite{liaw_tuneresearchplatform_2018} for hyperparameter tuning. All experiments are conducted at the Melbourne Research Cloud computation node with an AMD EPYC 9474F 48-core CPU and 128GB of RAM.

\subsection{Performance Study}
\subsubsection{QFOR Policy Training Performance}
To thoroughly evaluate the adaptability of QFOR across different operational priorities in quantum cloud environments, with the main priority to optimise the fidelity performance, we trained separate policy instances under different time weights $\beta \in \{0.5, 1.0\}$ and different hyperparameters to explore the orchestration trade-off and determine the balance configuration for optimizing the overall orchestration performance. We employed Ray Tune \cite{liaw_tuneresearchplatform_2018} for systematic hyperparameter optimization across all three $\beta$ configurations, evaluating key parameter combinations to identify optimal settings. The optimal hyperparameter configuration was selected based on convergence stability and final performance across all training modes, with the tuning result shown in Table~\ref{tab:hyperparameters}. Other hyperparameters and settings are based on the default configuration of PPO in Ray RLlib \cite{liang_rllibabstractionsdistributed_}.  


\begin{table}[htbp]
\centering
\caption{QFOR Training Hyperparameters}
\label{tab:hyperparameters}
\begin{tabular}{lc}
\hline
\textbf{QFOR Parameters} & \textbf{Value} \\
\hline
Learning Rate & $0.0001$ \\
Discount Factor ($\gamma$) & 0.9 \\
KL Coefficient & 1.0 \\
GAE Parameter ($\lambda$) & 0.95 \\
PPO Clip Parameter & 0.3 \\
Entropy Coefficient & 0.01 \\
Train Batch Size per Learner & 180 \\
Fidelity Reward Weight ($\alpha_1$, $\alpha_2, \alpha_3$) & 0.8, 0.1, 0.1 \\
\hline
\end{tabular}
\end{table}

\begin{figure}[htbp]
\centering
\hfill
\subfloat[$\beta = 0.5$\label{fig:training_beta05}]{
\includegraphics[width=0.45\linewidth]{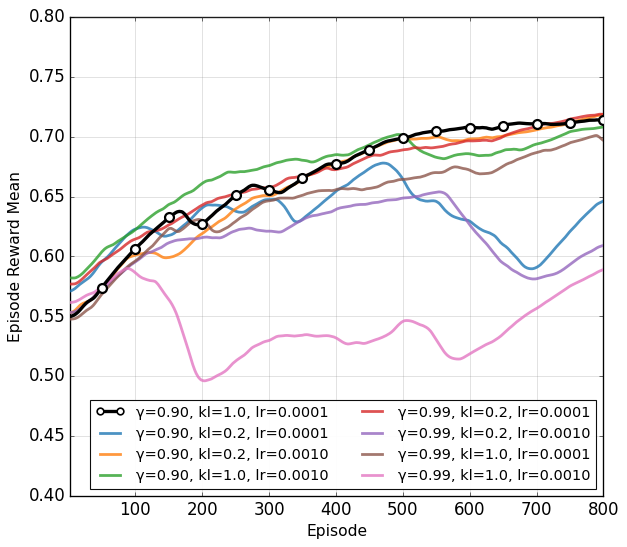}
}
\hfill
\subfloat[$\beta = 1.0$\label{fig:training_beta10}]{
\includegraphics[width=0.45\linewidth]{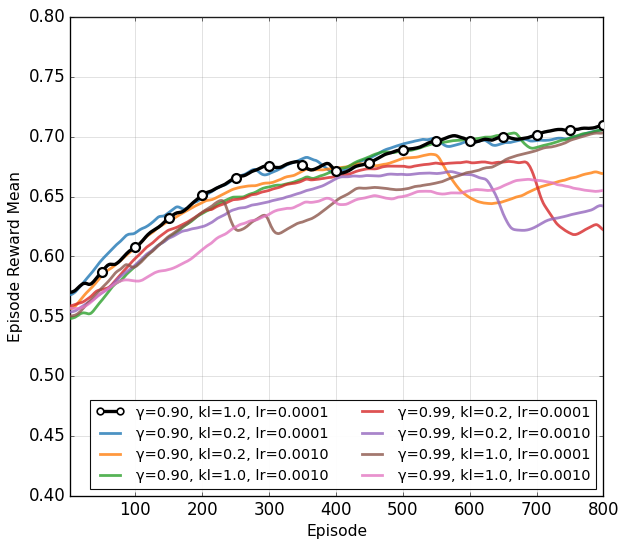}
}
\caption{Training convergence comparison across all configurations. The optimal hyperparameter configuration (black line with markers) achieves more consistent convergence in both time weight $\beta$ settings}
\label{fig:training_convergence}
\end{figure}

Figure~\ref{fig:training_convergence} shows training convergence across 800 training episodes. The optimal configuration across all configurations demonstrates superior performance with three key advantages: (1) stable reward convergence to approximately 0.70 by training episode 400-500 across all modes, (2) minimal variance indicating robust optimization dynamics, and (3) consistent performance across different $\beta$ values, demonstrating hyperparameter robustness. Other configurations exhibited training instability and performance degradation after episode 500-600, particularly in balanced and high-performance modes. The lower discount factor ($\gamma=0.9$) proves advantageous for quantum orchestration by appropriately balancing immediate scheduling decisions with long-term efficiency in dynamic environments.

\subsubsection{Execution Fidelity Performance Analysis}
    To evaluate the effectiveness of the QFOR policy, we conducted a comprehensive performance evaluation across 100 evaluation episodes with 6,000 quantum tasks using an independent test dataset distinct from the training data to ensure an unbiased assessment of the learned policies' generalization capabilities and practical applicability. We compared QFOR against four representative baseline policies, similar to the evaluation approach of other related works \cite{li_moiraioptimizingquantum_2024, nguyen_drlqdeepreinforcement_2024, luo_adaptivejobscheduling_2025}: 
\begin{itemize}
    \item \textit{Round Robin (RR)}: Distributes quantum tasks cyclically across nodes, ensuring fair resource allocation while potentially ignoring node-specific characteristics.
    \item \textit{Smallest Error First (SEF)}: Assigns QTask to QNode with the smallest average error rates of all gate operations.
    \item \textit{Fastest Duration First (FDF)}: Assigns QTask to QNode with the fastest average gate duration times.
    \item \textit{First Available Node (FAN)}: Assigns QTask to the first idle quantum node to minimize waiting time.
\end{itemize}

\begin{figure}[htbp]
\centering
\subfloat[Execution fidelity in all evaluation episodes\label{fig:fidelity_beta01}]{
\includegraphics[width=0.52\linewidth]{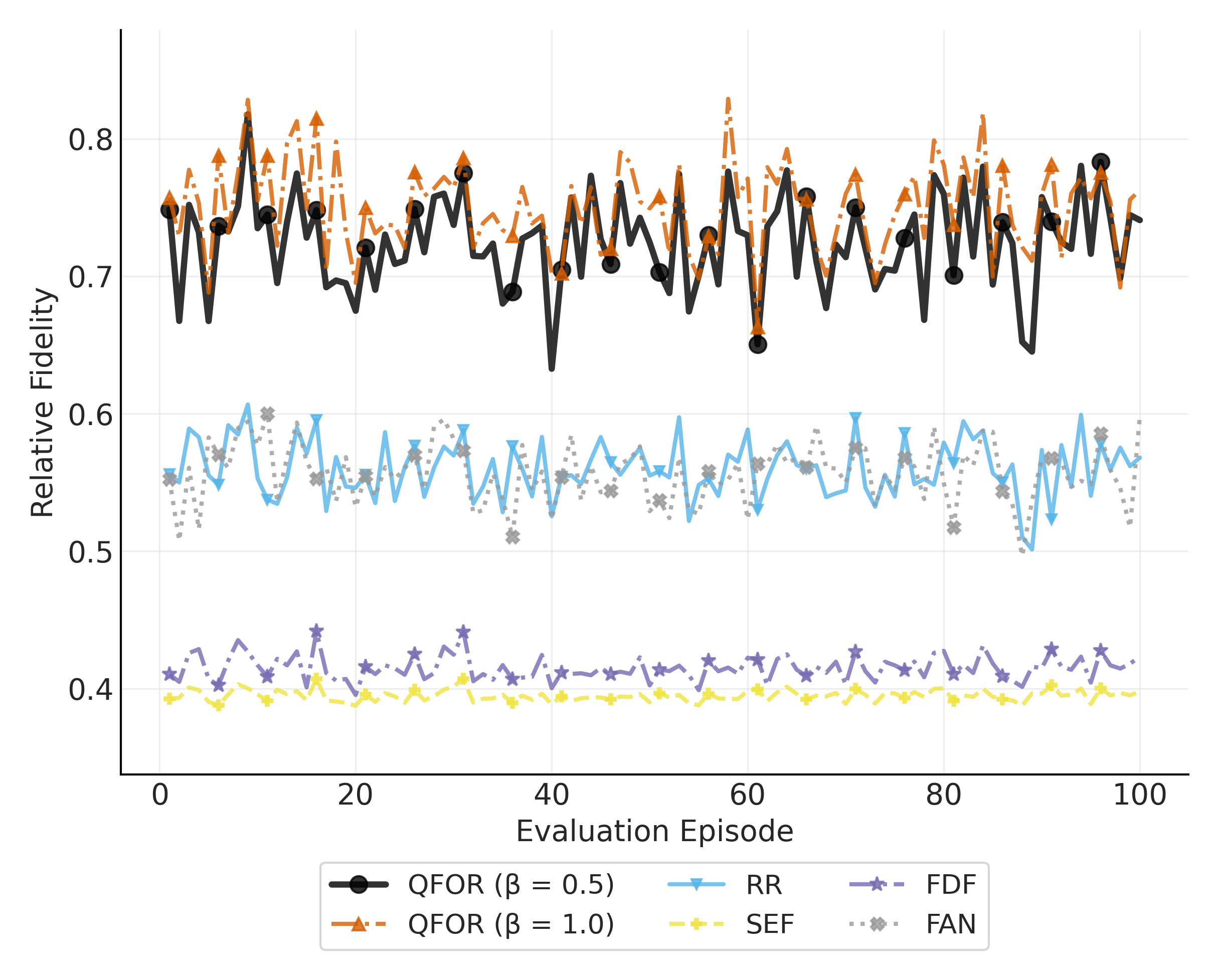}
}
\hfill
\subfloat[Average execution fidelity of all policies\label{fig:fidelity_beta05}]{
\includegraphics[width=0.43\linewidth]{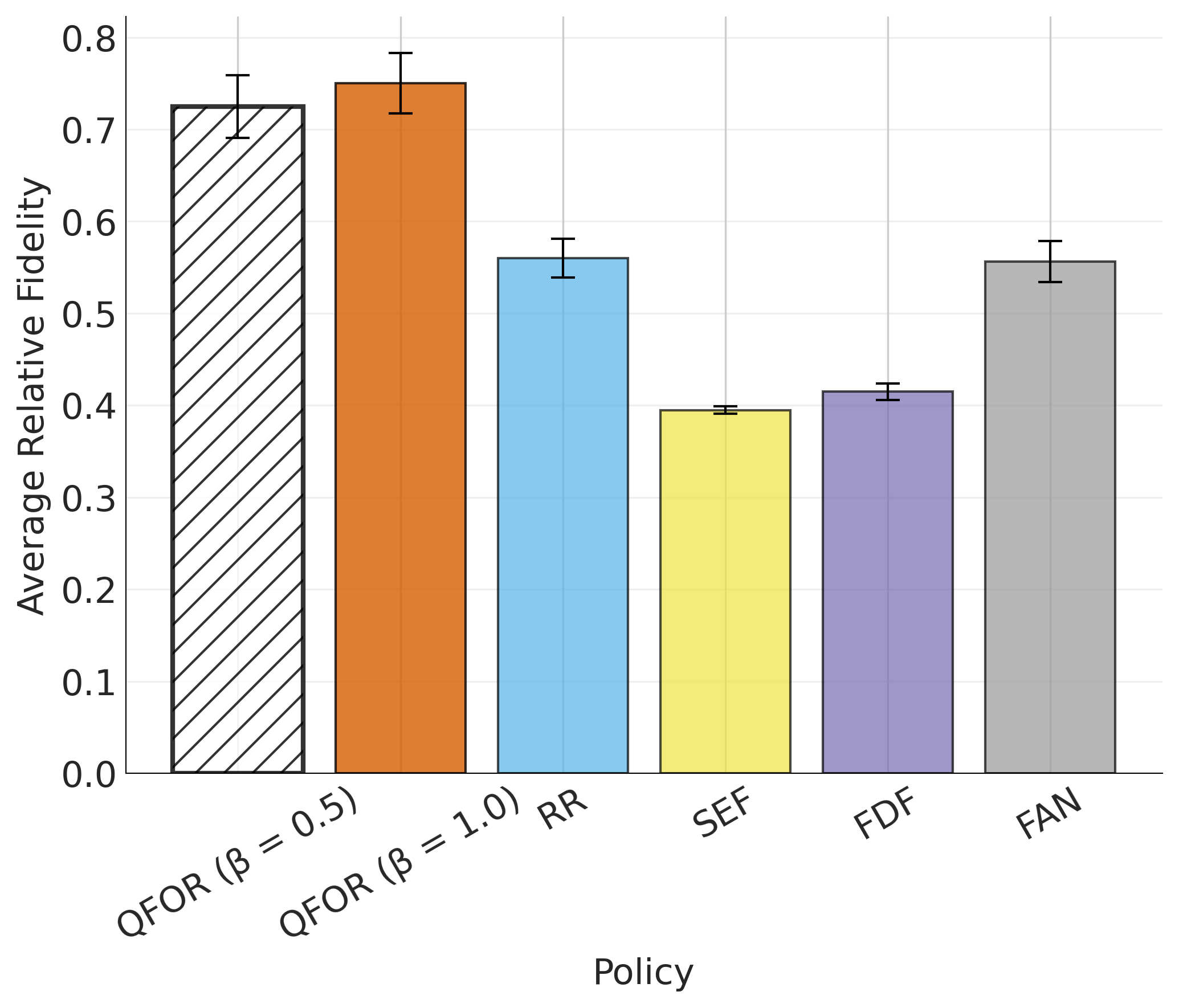}
}
\caption{Relative execution fidelity performance comparison across all policies over 100 evaluation episodes.}
\label{fig:fidelity_evaluation}
\end{figure}

These baselines represent the spectrum of conventional scheduling strategies commonly employed in distributed computing environments. Figure~\ref{fig:fidelity_evaluation} and Table \ref{tab:scheduling-performance} present the relative fidelity performance comparison across all policies. 

QFOR demonstrates substantial fidelity advantages across all operational configurations. The episode-wise analysis in Figure~\ref{fig:fidelity_evaluation}a shows QFOR's consistent enhancement throughout the evaluation period. Notably, all three QFOR configurations maintain stable performance in the 0.70-0.75 range, while baselines cluster in the 0.394-0.56 range.
As shown in Figure~\ref{fig:fidelity_evaluation}b, with error bars denoting standard deviation, QFOR achieves consistently higher relative fidelity performance, with an average fidelity score at 0.725  using $\beta = 0.5$, and 0.75 using $\beta = 1.0$. These results represent significant improvements over the best-performing baseline (RR at 0.56), with enhancement margins of 29.5\% and 34\%, respectively. In contrast, the worst-performing policy, SEF, which greedily selects the QNode with the smallest average error rate regardless of the scheduled quantum circuit structure, fails to maintain a high fidelity performance score, achieving a mean performance score of 0.395. This highlights that final execution fidelity depends not only on the average error rate of the QNode but also critically on the transpilation process, including the mapping of the QTask to the specific qubit topology of the QNode. These results indicate the key limitations of conventional scheduling approaches in quantum cloud environments and demonstrates robust policy learning that generalizes effectively to unseen quantum tasks and dynamic device conditions. The consistent high performance across different $\beta$ values demonstrates the ability of QFOR to learn context-aware scheduling policies that adapt optimization priorities while maintaining overall effectiveness. 




\begin{figure}[htbp]
\centering
\subfloat[Task completion time across evaluation episodes\label{fig:completion_average}]{
\includegraphics[width=0.48\linewidth]{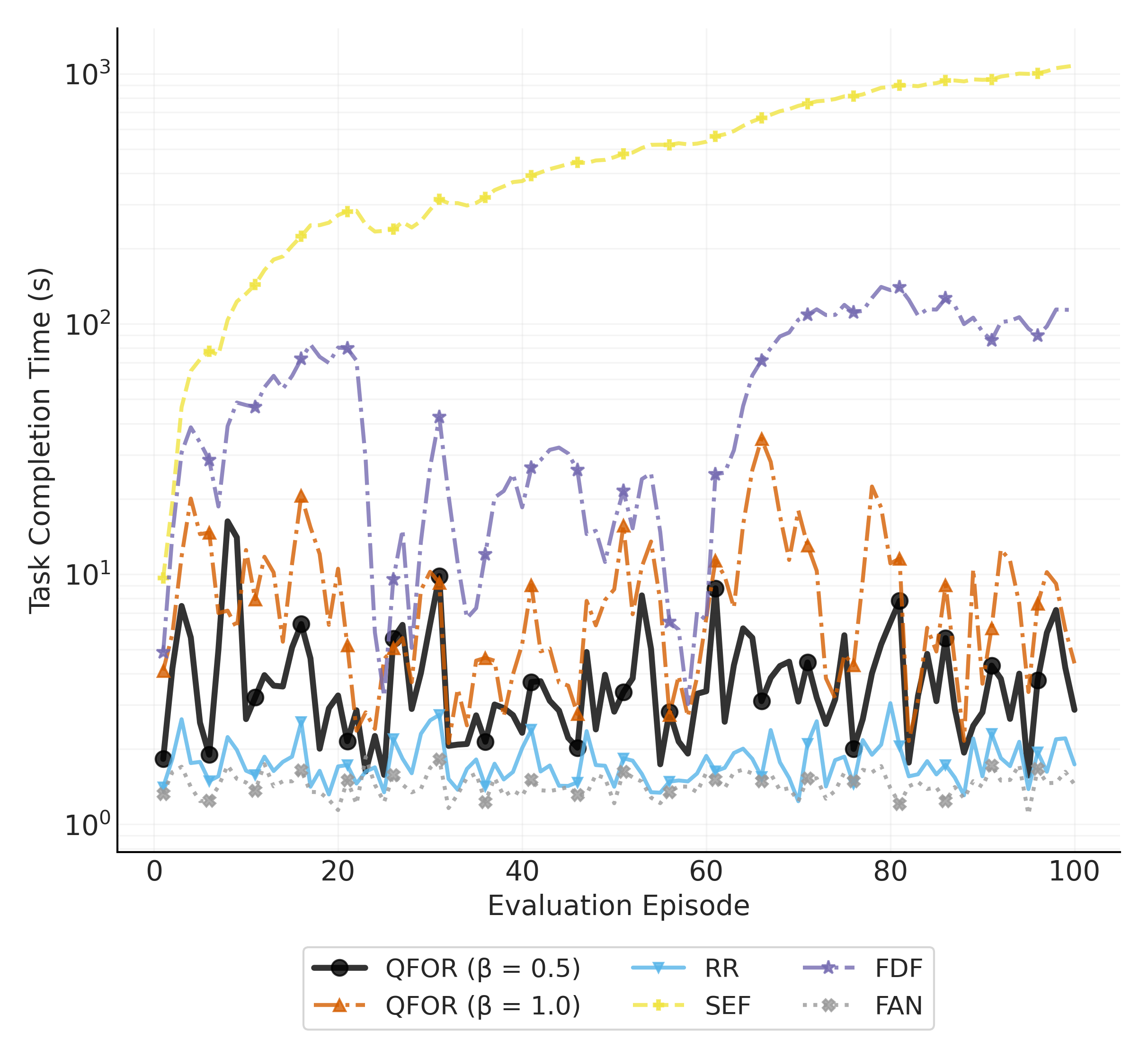}
}
\hfill
\subfloat[Average task completion time comparison\label{fig:completion_episodes}]{
\includegraphics[width=0.46\linewidth]{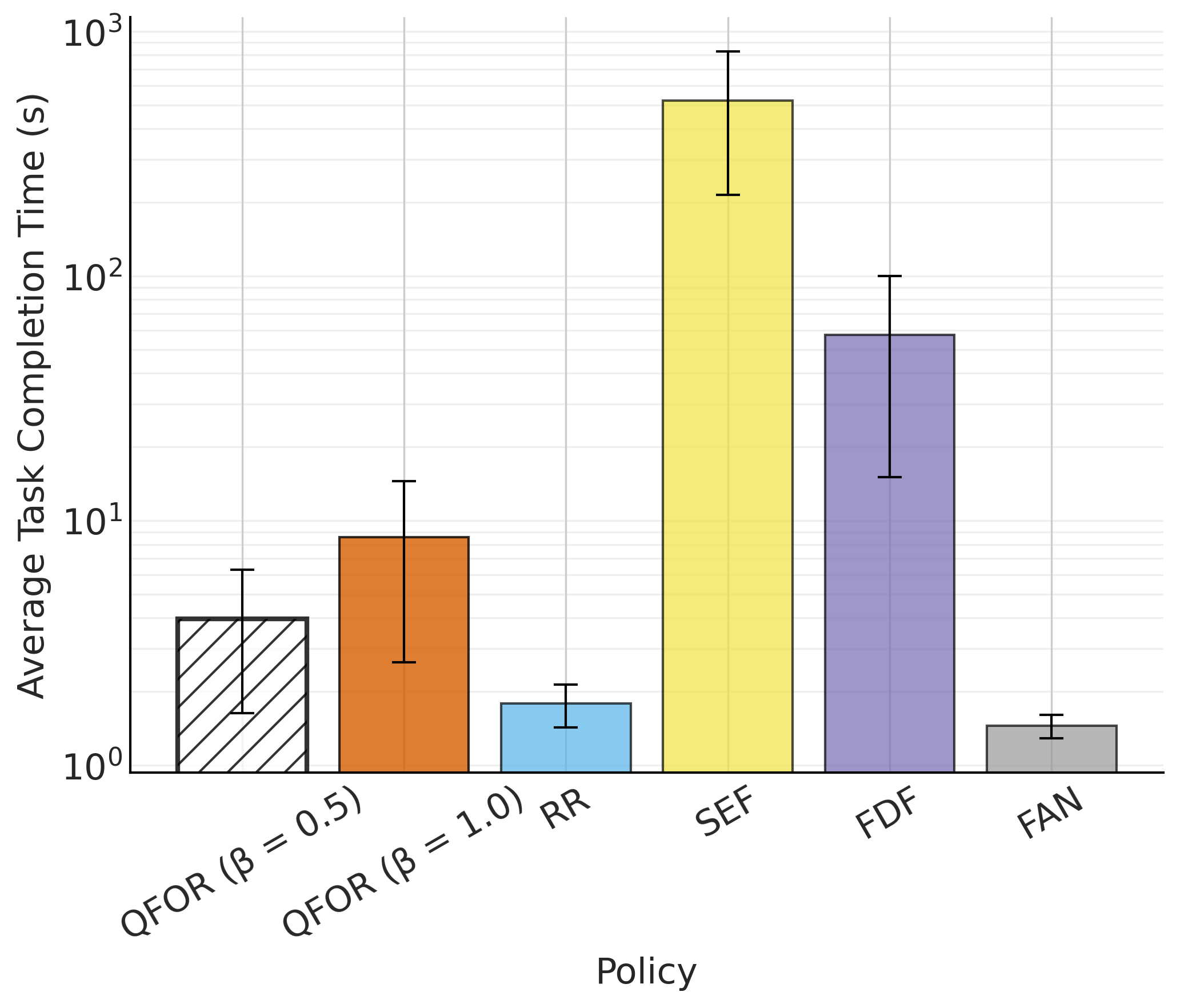}
}
\hfill
\subfloat[Task execution time across evaluation episodes\label{fig:completion_average}]{
\includegraphics[width=0.48\linewidth]{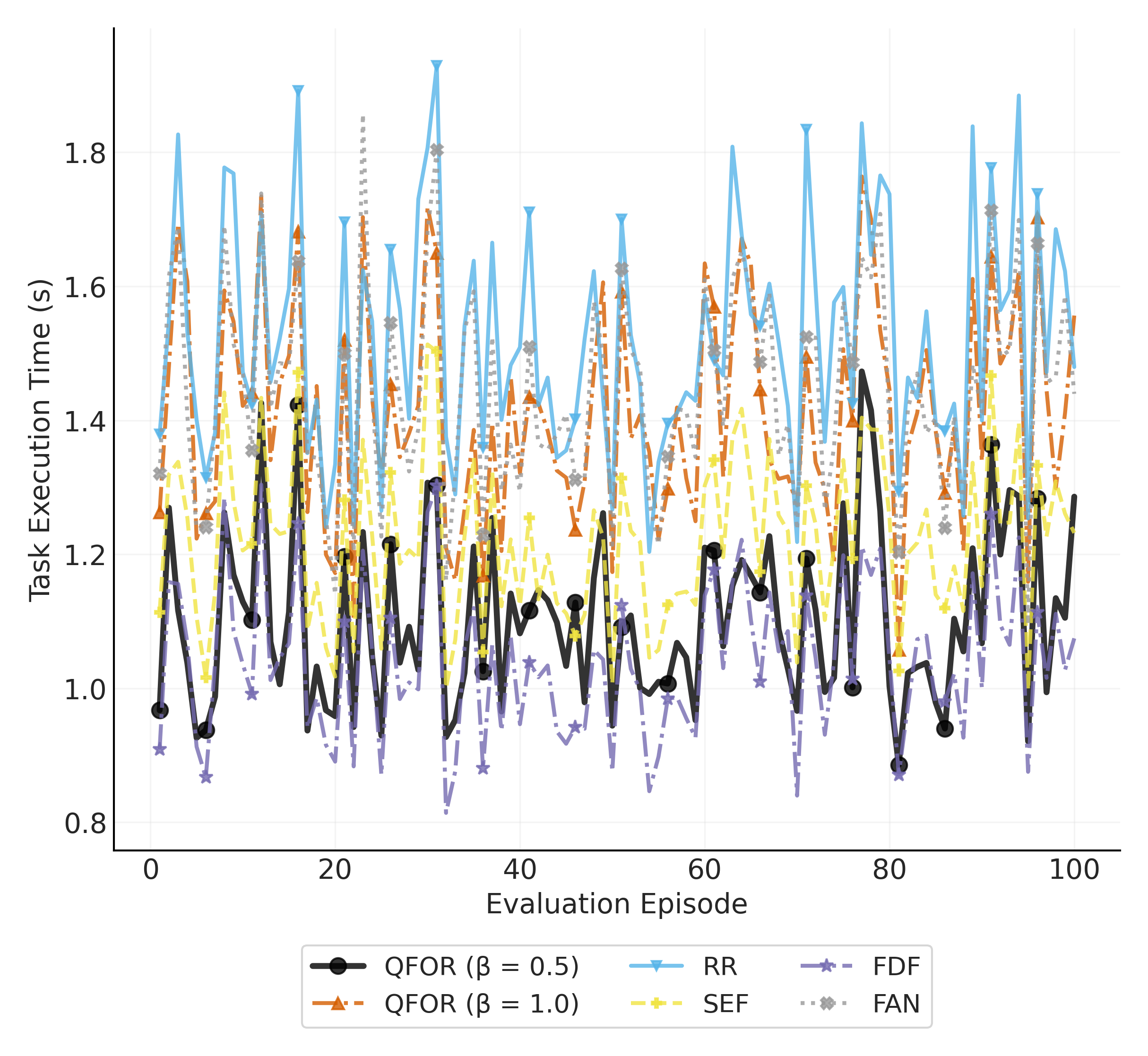}
}
\hfill
\subfloat[Average task execution time comparison\label{fig:completion_episodes}]{
\includegraphics[width=0.46\linewidth]{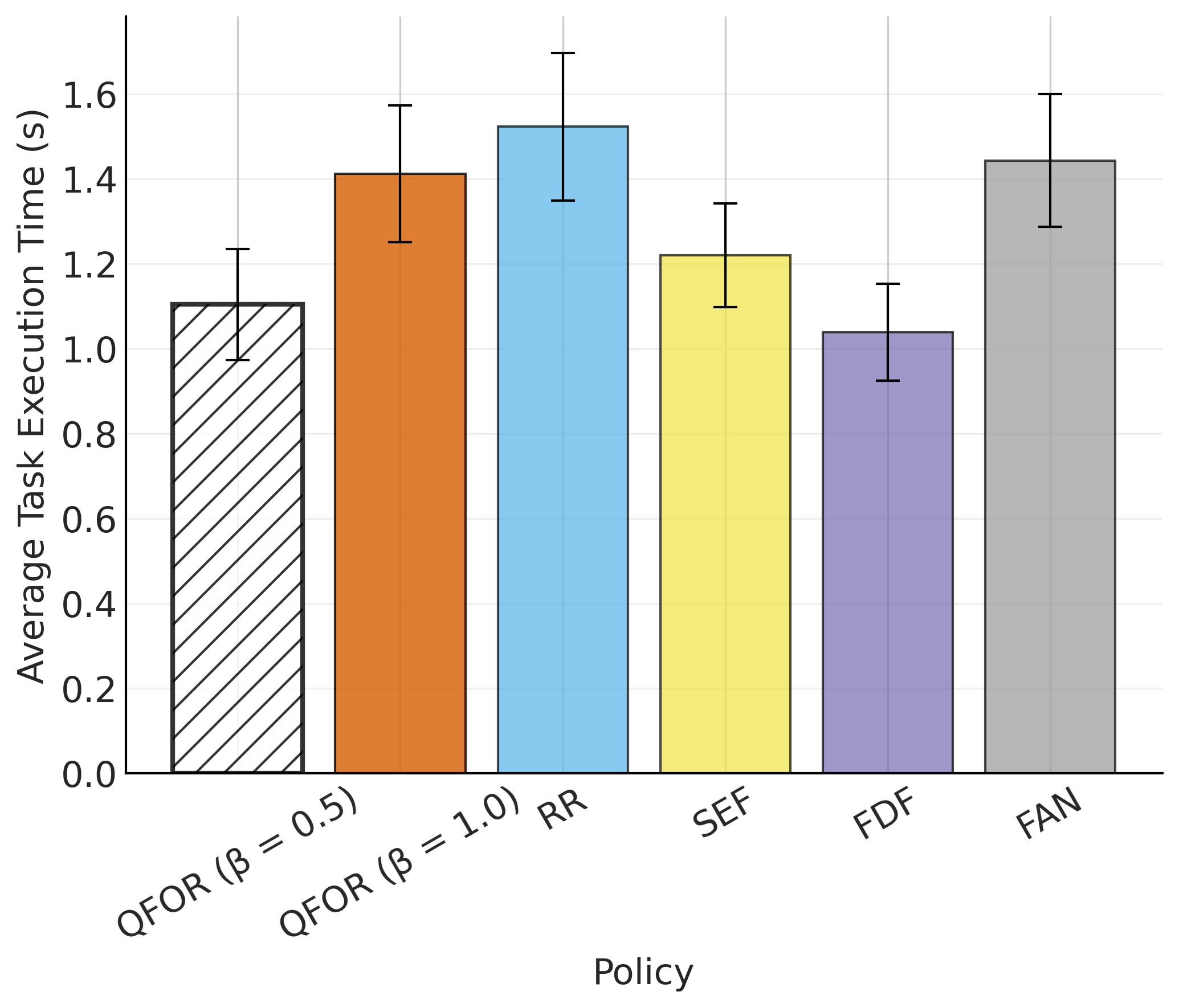}
}
\caption{Task completion and quantum execution time analysis of all policies over 100 evaluation episodes.}
\label{fig:completion_time_evaluation}
\end{figure}

\begin{table}[htbp]
\centering
\caption{Detailed performance comparison of all policies, regarding average relative fidelity score, execution time, and total completion time (± standard deviation) over 100 evaluation episodes}
\label{tab:scheduling-performance}
\begin{tabular}{lccc}
\toprule
\textbf{Policy} & \textbf{Average Fidelity Score} & \textbf{Average Execution Time (s)} & \textbf{Average Completion Time (s)} \\
\midrule
QFOR ($\beta = 0.5$) & $0.725 \pm 0.034$ & $1.104 \pm 0.131$ & $3.967 \pm 2.336$ \\
QFOR ($\beta = 1.0$) & $0.750 \pm 0.033$ & $1.412 \pm 0.161$ & $8.580 \pm 5.946$ \\
RR                    & $0.560 \pm 0.021$ & $1.523 \pm 0.174$ & $1.784 \pm 0.357$ \\
SEF                   & $0.395 \pm 0.004$ & $1.220 \pm 0.123$ & $521.675 \pm 306.631$ \\
FDF                   & $0.415 \pm 0.009$ & $1.039 \pm 0.114$ & $57.502 \pm 42.483$ \\
FAN                   & $0.556 \pm 0.022$ & $1.444 \pm 0.157$ & $1.448 \pm 0.159$ \\
\bottomrule
\end{tabular}
\end{table}

\subsubsection{Execution Fidelity-Time Trade-off Analysis}

As the main priority of the orchestration is optimizing the fidelity performance, with the secondary consideration being to balance the execution time required, we conducted a comprehensive analysis of total completion time and quantum execution time of all QTasks in the evaluation across all policies to find the optimal time weight to achieve this orchestration goal and explore the trade-off between fidelity and time in the decision. Figure~\ref{fig:completion_time_evaluation} and Table \ref{tab:scheduling-performance} present a detailed timing analysis across all policies, demonstrating both average performance and episode-wise progression over 100 evaluation episodes.

The results indicate that greedy policies focusing exclusively on gate error rates (SEF) or gate duration time (FDF) lead to substantial waiting times, resulting in significantly longer total completion times compared to other approaches. In contrast, policies such as First Available Node (FAN) and Round Robin (RR) effectively minimize waiting times by distributing tasks more evenly, thereby reducing overall completion time. 

As quantum execution time cost is critical and task queueing before execution can be handled by a classical controller, it is more valuable and cost-efficient to optimise the quantum execution time. Figures \ref{fig:completion_time_evaluation}c and \ref{fig:completion_time_evaluation}d show the actual quantum execution time of all policies. The result shows that the balanced time weight $\beta=0.5$ achieves a reasonable average execution time, which is slightly higher than the execution time greedy policy (FDF), while maintaining the fidelity score consistently higher than all of the other baselines, up to approximately 84\% higher relative fidelity performance score compared to the SEF baseline. This result suggests the optimal and balanced configurations of the time weight that encourage the QFOR policy to achieve high fidelity while maintaining the balance of fidelity-time tradeoff. Furthermore, minimising the execution time also inherently optimises the monetary cost of using quantum resources, which is essential in the current landscape of quantum cloud computing environments \cite{nguyen_quantumcloudcomputing_2024}. 

The results demonstrate that adaptive orchestration fundamentally balances the execution fidelity-time trade-off. While traditional approaches force a binary choice between speed and quality, QFOR policy identifies strategies that balance both objectives. This analysis establishes three key insights for quantum cloud orchestration: (1) Fidelity should be the key consideration in the current NISQ era as error sensitivity makes quality optimization essential as fast but inaccurate operations are ultimately ineffective, (2) Our adaptive, configurable policy enable flexible resource management optimization strategies which can be further extended, and (3) DRL-based is a potential approach that discovers non-obvious scheduling patterns that outperform conventional heuristics.

\section{Conclusion and Future Directions}
\label{sec:conclusions}

This paper presents QFOR, a novel fidelity-aware deep reinforcement learning framework for quantum task orchestration in heterogeneous cloud-based environments with NISQ computation resources. We developed a holistic orchestration technique specifically designed for optimizing overall fidelity performance, which is critical for quantum task execution. Our systematic approach integrates comprehensive quantum task execution and fidelity performance estimators based on device calibration data, enabling rigorous emulation of noisy quantum hardware behavior. The configurable orchestration objectives successfully balance execution fidelity and time across different operational priorities, with extensive evaluation demonstrating 29.5-84\% fidelity performance score improvements over traditional baseline methods. This work establishes the foundation and facilitates learning-driven quantum resource management research in hybrid quantum-HPC systems, which require additional effort to develop a robust and adaptive framework and techniques to keep up with the advances of quantum hardware development.

While QFOR demonstrates potential in quantum cloud orchestration, we recognized several limitations that present opportunities for future research. First, dynamic device calibration data can be considered to mimic the continuously updated device characteristics and adapt scheduling policies in real-time. We plan to explore the investigation of QFOR performance on large-scale quantum cloud infrastructures with dozens of quantum devices and circuits requiring a larger number of qubits. 
We also plan to incorporate more rigorous approaches to multi-objective optimization in future work. Furthermore, circuit knitting \cite{piveteau_circuitknittingclassical_2024} for distributing large quantum tasks across distributed quantum systems can be enhanced by utilizing realistic quantum circuit simulations instead of synthetic data, making it more suitable for practical environments. Additionally, the centralized policy learning architecture may become a bottleneck in distributed quantum computing environments, motivating future exploration of distributed reinforcement learning architectures \cite{qiu_distributedcollectivedeep_2021, goudarzi_distributeddeepreinforcement_2021} where multiple agents coordinate scheduling decisions across geographically distributed quantum resources. These future directions will be essential for realizing the full potential and optimizing the resource management of quantum cloud computing as the technology matures toward practical quantum advantage applications.

\section*{Acknowledgments}
Hoa Nguyen acknowledges the support from the Science and Technology Scholarship Program for Overseas Study for Doctoral Degrees, Vingroup, Vietnam. 


\bibliographystyle{ACM-Reference-Format}
\bibliography{references}

\end{document}